\documentclass[a4paper, 10pt, conference]{ieeeconf}      

\IEEEoverridecommandlockouts                              
\usepackage{hyperref}
\usepackage[pdftex]{graphicx}

\title{\LARGE \bf
A privacy-preserving system for data ownership using blockchain and distributed databases
}

\author{Sabine Bertram$^{*}$ and Co-Pierre Georg$^{*}$
\thanks{$^{*}$Both authors are with the Financial Innovation Lab at the University of Cape Town. Email: brtsab001@myuct.ac.za; co-pierre.georg@uct.ac.za}%
\thanks{First version: August 7, 2018. This version: \today.}%
}

\begin{document}

\maketitle
\thispagestyle{empty}
\pagestyle{empty}

\begin{abstract}

Blockchain has the potential to revolutionize the way we store, use, and process data. Information on most blockchains can be viewed by every node hosting the blockchain, which means that most blockchains cannot handle private data.
Decentralized databases exist that guarantee privacy by encrypting user data with the user's private key, but this prevents easy data sharing.
However, in many real world applications, from student data to medical records, it is desirable that user data is anonymously searchable.
In this paper we present a novel system that gives users \emph{ownership} over their data while at the same time enabling them to make their data searchable within previously agreed upon limits.  Our system implements a strong notion of ownership using a self-sovereign identity system and a weak notion of ownership using multiple centralized databases together with a blockchain and a tumbling process.
We discuss applications of our methods to university's student records and medical data.
\end{abstract}

%
\section{INTRODUCTION}
%

The Stanford Encyclopedia of Philosophy defines ownership and private property as ``a kind of system that allocates particular objects [...] to particular individuals to use and manage as they please, to the exclusion of others [...] and to the exclusion also of any detailed control by society.'' \cite{SEP2016.Ownership}.
The key word here is \textit{exclusion}. In the context of data, ownership implies that the owner of a datum can exclude others from access or use. 

The web 2.0 was built on the premise of user-generated data \cite{OReilly2005.Web20}. However, under the current paradigm, users have no ownership of the data they create. To the contrary, the business model of companies like Google, Facebook, and Twitter is to offer free services financed by selling user data to advertisers and other third parties.

In this paper, we describe a system which implements the notion of \emph{data ownership} using a decentralized database, multiple centralized databases administered by different entities and a public blockchain as a link between them. Our system allows data creators to store de-identified user data, such as student records or medical data, in a decentralized database. Identifying information is stored in several centralized databases, that allow updating of data without appending it, managed by custodians. Alternatively, identifying information can be stored decentralized using self-sovereign identity (SSI) system.
User data is linked to identifying information using smart contracts on a blockchain. Crucially, users obtain ownership of their data by \emph{scrambling} the link between user data and identifying information on the blockchain. This way, a user can exclude others from using her personal information by preventing them from linking anonymous user data with identifying information. 

The immutable append-only nature of blockchains makes them fundamentally different from traditional databases where a client can also perform a delete function in addition to the create, read, and update function that a blockchain performs as well \cite{Martin1980.CRUD}.

There are three major drawbacks when using a blockchain.
First, blockchains are auditable by design, which is in stark contrast to the need for greater privacy highlighted in the recent NSA surveillance\footnote{See \href{https://www.nytimes.com/2013/06/08/technology/tech-companies-bristling-concede-to-government-surveillance-efforts.html}{NY Times} (accessed 08 July 2018).} and Cambridge Analytica\footnote{See \href{https://www.theguardian.com/uk-news/2018/may/02/cambridge-analytica-closing-down-after-facebook-row-reports-say}{The Guardian} (accessed 08 July 2018).} scandals.
Second, blockchains do not scale well \cite{Croman2016.Scalability}. Some estimate that about 90\% of the world's data was generated in the last few years alone.\footnote{See \href{https://www.bbc.com/news/business-26383058}{BBC} (accessed 08 July 2018).} Thus, blockchains are not a viable alternative for most real-world applications.
And third, searching on a blockchain is difficult. While public blockchains are fully auditable, searching by the value of a transaction rather than the address (i.e. the key at which the transaction can be found on the blockchain) is cumbersome. It is easy to get a list of all transactions recorded on the blockchain, but finding all transactions of a certain size usually requires querying the blockchain, which is slow and expensive.

One solution to the privacy challenges outlined above is self-sovereign identity \cite{Birch2014.Identity,TobinReed2016.Sovrin}.
A number of projects aim to solve the privacy issue of blockchains by introducing self-sovereign identity \cite{Lundkvist2016.uPort,LinghamSmith2018.Civic}. These approaches have in common that they store user's private keys offline and thus decentralized, usually in a mobile application that signs and validates data e.g. for interactions with a blockchain. Additionally, it can store any kind of private data offline on the mobile device. Self-sovereign identity solves the problem that private keys cannot be safely stored on a massive centralized database since this database would then be an extremely attractive target for attackers (as argued, for example, in \cite{Lundkvist2016.uPort}).

There are two problems with self-sovereign data, though. First, among legal scholars it is subject to debate whether complete data ownership is a goal that lawmakers should pursue in the first place \cite{Evans2011.MuchAdo}. De-identified medical data, for example, can be a public good if it is used for research purposes. So one could make an argument that this data should be owned by the government, or that there should be strict limits under which individuals can own their data. But even if these legal---and in some instances, moral---issues can be resolved, a second, technological, problem remains. If a user encrypts her data with her private key before storing it in a database, this data cannot be searched over. So any query on \emph{another} datum that involves relative measures will remain incomplete. For student data it might be relevant to know not only the mark of a student, but whether she was one of the top 5\% achiever in a given year. For medical data the importance of relative measures is even clearer. Whether the results of a blood test show an anomaly is by definition a relative measure, for example.

Self-sovereign identity systems can guarantee that only the owner of a datum can access it. But they also prevent any use of the data by third parties, even in cases where the owner would allow third-party access to the de-identified data.
Ideally, the data owner would be able to write a contract that prevents the third party from copying and selling the user data. This is not possible, however, since the third party cannot commit to this agreement. The consequence of this commitment issue is that users are unwilling to share their data at all and the market breaks down.
Another major issue is the lack of adoption of self-sovereign identity systems. While various projects exist to promote self-sovereign identities, it will require time before users adapt. Therefore, any system that depends on self-sovereign identity adoption will inevitably face difficulties in scaling. Our system enables a weak notion of data ownership even \textit{without} a self-sovereign identity system.

Our privacy-preserving system for data ownership solves both the legal and technological issues outlined above. A custodian (e.g. a university or a hospital) creates user data which is stored on a decentralized database. This allows other custodians to share the same database without being able to identify individuals and has the added advantage that the authenticity of the data is never in question because the data is cryptographically signed by the creator. De-identified user data is linked to identifying data via a smart contract on the blockchain that stores a user's id in the identifying database. If a user wants to claim ownership of the data, she has to break the link between the smart contract and the identifying database her data is stored in by either updating her id in the identifying database but not in the smart contract, or by moving her identifying data--and only her identifying data--onto her self-sovereign identity system.

Several blockchain-based projects want to store private data on the blockchain, including \href{https://medicalchain.com}{Medicalchain}, \href{https://www.bcdiploma.com}{BCDiploma}, and \href{https://aeron.aero/}{Aeron}. However, what all these projects have in common, and what sets our system apart, is that they do not implement data ownership, privacy, and searchability of data at the same time. Our system achieves all three goals--highlighted in Section \ref{Sec::GoalsLimitations}--simultaneously and we outline the system architecture in Section \ref{Sec::SystemArchitecture}. In Section \ref{Sec::UseCases} we discuss two possible use cases, while Section \ref{Sec::Conclusion} concludes.

%
\section{GOALS AND LIMITATIONS}\label{Sec::GoalsLimitations}
%

Before turning to the specification of our system, it is useful to re-iterate the three goals our system achieves:
\begin{enumerate}
	\item[(i)] Implement a notion of data ownership which allows users to exclude others from using their data.
    \item[(ii)] Allow third parties to compute previously agreed upon operations on de-identified user data.
    \item[(iii)] Store the user data of possibly millions of users, i.e. operate at scale.
\end{enumerate}
Our system consists of three components, a decentralized, peer-to-peer database, multiple centralized databases which allow for the deletion of entries, and a blockchain to link those two components. Arguably, the biggest advantage of our system is its simplicity. Our main goal was to achieve a notion of data ownership while still allowing third parties access to de-identified data under previously agreed upon conditions.

\subsection{Our notion of ownership}\label{Sec::SystemArchitecture:Ownership}
No universally accepted notion exists of what it means to \emph{own} data. Even the legal framework covering data protection varies widely across countries. In 2016 the European Commission, for example, released a report studying data protection and ownership across the European Union \cite{EC2016_DataOwnership}. While countries like Germany confer \textit{``ownership-like rights on a data holder''}, there is \textit{``no statutory basis for the protection of data as such, and so no property rights subsist; data cannot be stolen, assigned, or inherited.''} in the United Kingdom. In the United States, the US Department of Health and Human Services is responsible for guidelines on medical data. In an article on data ownership, it highlights the importance of data access and, following \cite{Loshin2002.DataOwnership}, \href{https://ori.hhs.gov/education/products/n_illinois_u/datamanagement/dotopic.html}{states} that: \emph{``The control of information includes not just the ability to access, create, modify, package, derive benefit from, sell or remove data, but also the right to assign these access privileges to others.''}

These examples are not supposed to give an exhaustive overview of the legal debate, but rather highlight the range of opinions expressed in this debate. The main trade-off faced by lawmakers is between data owners' privacy and the use of data in the creation of public goods such as medical research. Full ownership implies that the owner can exclude others from using the data. But in many applications, this right to privacy needs to be carefully weighed against the contribution to a public good. Or, as the European Commission puts it in their report \cite{EC2016_DataOwnership} \emph{``[$\ldots$] a regime, which incentivises stakeholders to withhold access for third parties to their data altogether, potentially restricts the extent of exploitation and innovation within data driven industry sectors.''}

Our approach to ownership loosely follows the economic literature, and in particular the literature on incomplete contracts and the notion of ownership in contract theory \cite{GrossmanHart1986.TheoryFirm}. Here, contractual rights are either specific rights or residual rights and ownership is the control over these residual rights. The distinction between specific and residual rights becomes important when two parties form a contract about the use of an asset. To make things concrete, consider a user selling data to a company. Because data can be copied at almost no cost and with only very limited copyright protection (an exception is the EU General Data Protection Regulation \cite{EU2016.GDPR}), the user has almost no control over what happens with her data once she sold them--she cannot exclude others from using her data and has thus no ownership over them. She also has no control over the residual rights, i.e. all rights to the data that are not specified in the original sales contract.

We say a user \emph{owns her data if she can prevent others from associating her user data with identifying information.} While in some instances this definition is much weaker than the definitions discussed above, in other instances the two are equivalent. Specifically, when the user data is valuable \emph{only} in conjunction with identifying information the ability to prevent others from associating user data and identifying information is equivalent to the ability to exclude others from using the data. For the situations we are concerned with, accessing student or medical data, our notion of ownership is sufficiently strong. In the case of student data, a potential employer, for example, will not be interested in the marks of an anonymous student, but rather in hiring a specific student.

\subsection{Limitations of our system}\label{Sec::SystemArchitecture:Limitations}
We built our system with a very specific goal in mind: to implement the notion of ownership for sensitive personal data described in \ref{Sec::SystemArchitecture:Ownership}. Our notion of ownership is a relatively weak one, though, since our system only prevents the mapping from de-identified user data from identifying information. In some instances it might be possible to reconstruct a user's identity from her user data \cite{Culnane2017.Re-identification,Narayanan2008.Re-identification}. The likelihood of this increases the more user specific data is stored in the same distributed database \cite{Narayanan2008.Re-identification}. However, the nodes themselves are run by the custodians, which requires a minimum level of trust (in the custodians), so our system is not trustless. The API endpoints are accessible only to approved third parties, which limits the potential for misuse.

Consequently, our system works well when user data cannot be used to substitute the identifying information. For the two use cases we have in mind, student data and medical records, this is likely to be the case. Student data is valuable, for example, to possible employers and companies offering bursaries and loans. But most of its value comes from the combination with identifying information since employers ultimately want to contact the student. Similarly, in the medical field most of the value of patient data comes when e.g. insurance companies knows who the patient is.

The second limitation of our system is the use of multiple centralized databases which allow for the deletion of data. Few distributed databases exist that provide that feature and they are not well developed yet\footnote{\href{https://tiesdb.com/}{TiesDB} is still in alpha mode and supposed to be a public platform where data storage is paid for. However, our legal requirements ask for a permissioned distributed database that is run by a consortium of custodians without additional costs. The same holds for \href{https://bluzelle.com/}{Bluezelle}. An alteration of both technology's code basis may be possible to run it as a private consortium, similar to how Quorum was forked from Ethereum as a private consortium.}. This concept is in stark contrast to the conventional blockchain ideology where the blockchain acts as a distributed ledger that keeps track of the entire history of all transactions, i.e. is append only. We explicitly allow for the possibility of self-sovereign identity data hosted by users through the use of an according system. But especially during a transitional period there will be users who do not want to use a self-sovereign identity system and rather prefer a centralized point of access to their identity data. This is very similar to how cryptocurrency exchanges like \href{http://www.coinbase.com}{coinbase} store their users' private keys.

We achieve security and transparency through a dedicated protocol which includes making the source code of our system open source. The single point of failure where users' identifying data is stored in centralized databases is still not ideal. However, as self-sovereign identity becomes more widely adopted, the need for this single point of failure will disappear and over time the system will become fully decentralized.

%
\section{SYSTEM ARCHITECTURE}\label{Sec::SystemArchitecture}
%

Our system consists of three principal components: (i) a decentralized and distributed database for user data; (ii) a number of centralized CRUD databases for identifying information; and (iii) a blockchain storing information how the two are linked. In the long run we expect the identifying information to be stored on an open self-sovereign identity system which would establish a strong notion of ownership. Until then, our system implements a weak notion of ownership by using a CRUD database together with a tumbling mechanism outlined below.

The decentralized and distributed record database stores user data. The database is hosted by a group of custodians (e.g. a university or a medical professional, depending on the application). Custodians have write access to this database and can query it through a third party interface. Identifying information about the data owners is stored only in the identifying databases or, prospectively, in a self-sovereign identity system. The link between user data and identifying information is stored in a smart contract on a blockchain owned and controlled by the data owner. The data owner can break the link by tumbling the entry in the coorsponding identifying database: generate a new random id for her personal information in the identifying database, update the personal information, but \emph{not} update the id stored in the smart contract. Then, the link between user data and identifying information is broken and from this moment forward her user data can no longer be matched with identifying information, which is how the data owner can exclude third parties from knowing about her identity.

The alternative is that the data owner uses a self-sovereign identity system, which ensures that every query to her identifying information has to be approved. Third parties can use a web interface to query user data with the permission of the custodian. The custodian decides which queries he allows on the user data through the API endpoints of the distributed user database. So only queries approved by the custodian can be executed, which is how the custodian keeps full control over the data he creates. User data is tamper proof, since only custodians have write access, while data owners and third parties only have read access.

\begin{figure*}
	\centering
	\includegraphics[scale=.45]{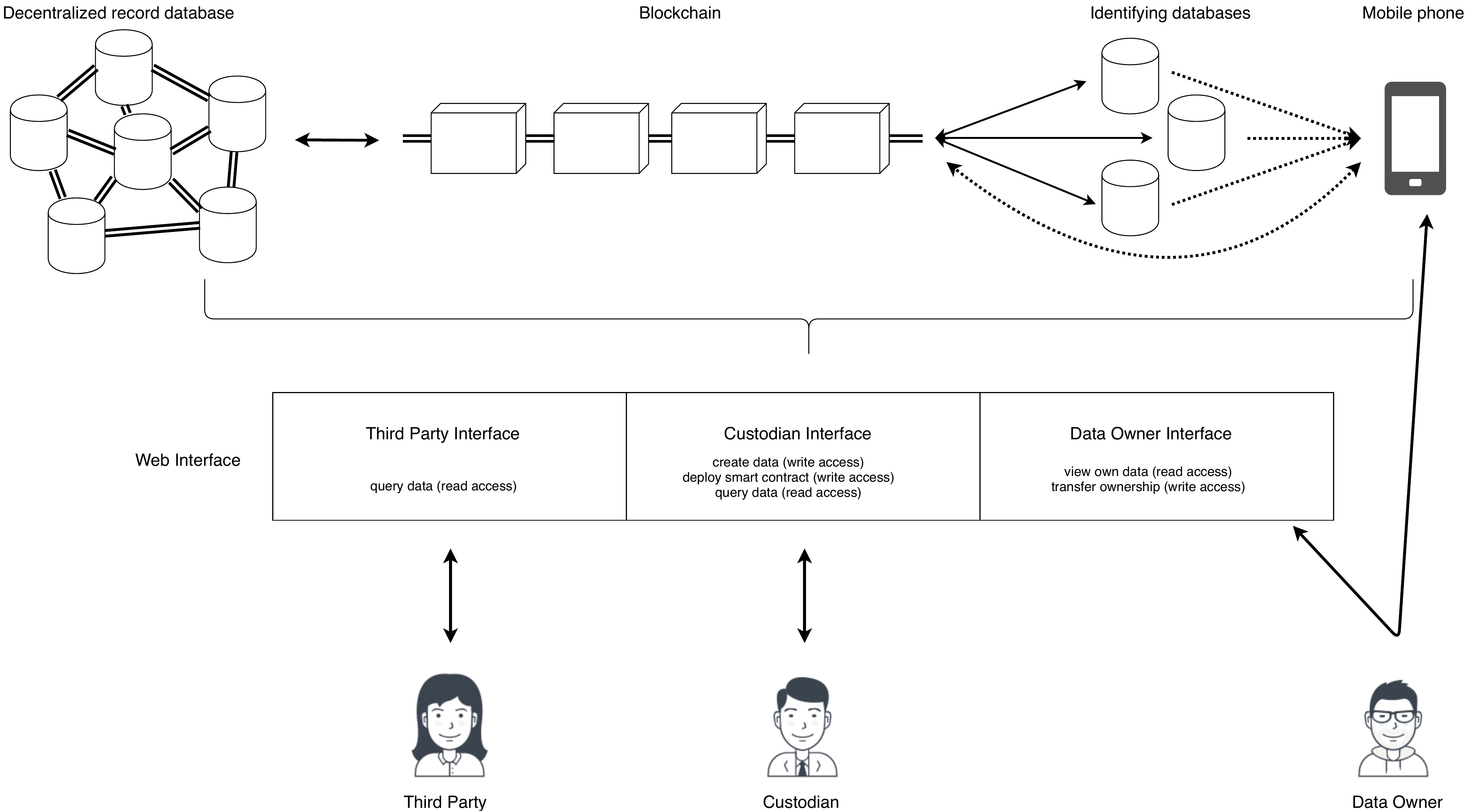}
	\caption{\textit{System architecture. A decentralized record database (left) stores user data. An entry in this database is linked with an entry in an identifying database (center right) or a self-sovereign identity stored on a mobile phone (right) through a smart contract hosted on a blockchain (center). Custodians host the decentralized record database and a identifying database and create data entries. Data owners control the smart contract and can break the link between user data and identifying information. Third parties can use a web interface to query user data with the permission of the custodian. Third parties can also query matching identifying information for a user datum, but only with the permission of the data owner.}}\label{Fig::Architecture}
\end{figure*}

The system is shown in Figure \ref{Fig::Architecture} and each component is explained in greater detail below. The tumbling process and CRUD databases for identifying information ensure at least a weak notion of ownership while data owners transition to a self-sovereign identity.\footnote{Especially in an emerging markets context this component of the system is important, as there are students who do not have their own cell phones and therefore could not use a self-sovereign identity system.}

\subsection{Storing user data using a decentralized and distributed database}
Our system follows a clear separation of concerns. User data is stored in a decentralized and distributed database. This database is the core data storage, holding all data of interest except identifying information. Ideally, this decentralized database is hosted by the custodians of the user data. These custodians form a consortium because they have a shared interest in the integrity of the user data. In the case of student data, universities are the custodians of the data and thus ideally suited to host the user database. Note that the requirements in terms of load, disk space, and bandwidth to operate the decentralized database increase in the number of members of the consortium and thus the number of users on the system. However, since the number of users is bound by the number of students, which is not too large, these requirements are moderate.


By storing the address of the corresponding smart contract with each datum, identification of user data is only possible if it is permissioned by the data owner (i.e. if the smart contract contains the link to the identifying information). User data can be created by the custodians only which ensures the integrity of the data at all times.

It is important to note that user data is \emph{not} encrypted in the decentralized database, but stored in plain text. This storage concept allows for queryability and searchability without revealing personal information. Two aspects of our setup alleviate potential privacy concerns with storing user data unencrypted. First, we carefully split user data and identifying information, so that no personal information (e.g. what marks a specific student has, or which medications a specific patient receives) can ever be revealed. And second, we only allow access to this data through an API which implements a notion of role based access control. The remaining concern is that database administrators at one of the custodians could potentially access the raw data. This is true, but the only data they would have access to is de-identified information which has little value without the identifying information.

\subsection{Storing identifying information in centralized CRUD databases}
One of the biggest problems of self-sovereign identity systems is user adoption. For many users, these systems are still too difficult to use and as a result, take-up is slow. The guiding principle of our system is to provide a notion of ownership to \emph{all} users, including those who do not want to use a self-sovereign identity system. Therefore, initially, all identifying information is stored in centralized CRUD databases, one hosted by each custodian.

A datum is hashed on the individual level, including a time stamp and a nonce, to create a unique id. To increase anonymity by obscurity, a certain percentage of fake data entries is created.
Depending on the degree of ownership the user wants to enforce, her entry's id is either updated in a smart contract on the blockchain owned by the user, or removed alltogether. Applying the weak form of ownership will update the nonce, thereby creating a completely different id hash, which is not updated in the smart contract, hence making it impossible to identify the user and rendering the user data worthless.

To decrease the chance of mapping this user to her data, a number of fake data entries' ids are also updated. 
To enhance the security of the system, user data and identifying information will be run on two physically different systems by the custodian. Alternatively, the identifying information databases can be hosted by multiple third parties. Security is further improved by disclosing all source code that interacts with the databases and by hiring an independent security audit company to check the source code for possible vulnerabilities.

The data is created by custodians only. However, the data owner is allowed to update the id or move the entry to her self-sovereign identity management system, thereby claiming ownership of the data.

\subsection{Linking user data and identifying information using the blockchain}
To link identifying information to user data and to ensure ownership and privacy, we use  blockchain technology. Individual-specific smart contracts store the id of the database entry containing the identifying information, i.e. the hash of that information plus nonce, and the link to the specific identifying database API endpoint that includes that datum. Furthermore, they include a flag that determines whether a third party is allowed to access their personal information. On creation, these smart contracts will be owned by the data creators, i.e. the custodians. Individuals are encouraged to claim ownership of their data and update their data id frequently to increase security. If an individual decides to revoke access to her data, she can update her id in the identifying database but not update it in the smart contract. 

To facilitate the search of smart contracts by a user's identifying id, one query contract is deployed that stores a mapping of id to smart contract. The value is set to 0 if the user claims ownership.

\subsection{The self-sovereign identity system}
To facilitate data ownership, we make use of an self-sovereign identity system. If a user decides to claim ownership of her data, she can choose between two different forms--weak and strong ownership. Weak ownership will update the user's id in the identifying database without updating it on the blockchain. Strong ownership will delete the user's entry from the identifying database and store it on the user's mobile device. In both cases, the user will become the owner of the smart contract on the blockchain. The process is made accessible through the help of an self-sovereign identity system that manages the user's key pair and the data storage, if requested. Obviously, strong ownership is the more desired form of ownership because the user can decide on a case-to-case basis whether she wants to reveal her personal data. Additionally, there is no single point of failure in form of a centralized identifying database anymore.

\subsection{The first implementation}
We will briefly focus on the actual technology we used to implement the first prototype of the system as it has been described above. The decentralized and distributed database that stores user data is a \href{https://www.bigchaindb.com/}{BigchainDB}, the centralized databases, storing identifying information, are \href{https://www.mongodb.com/}{MongoDB}s, and the blockchain is a private \href{https://ethereum.org/}{Ethereum} node. Self-sovereign identitiy is facilitated by \href{https://www.uport.me/}{uPort}. The BigchainDB, MongoDBs and the Ethereum blockchain are wrapped in APIs that deal authentication and authorization of the system. These APIs can be audited due to their open-source nature.

%
\section{USE CASES}\label{Sec::UseCases}
%

\begin{figure*}
	\centering
	\includegraphics[scale=.45]{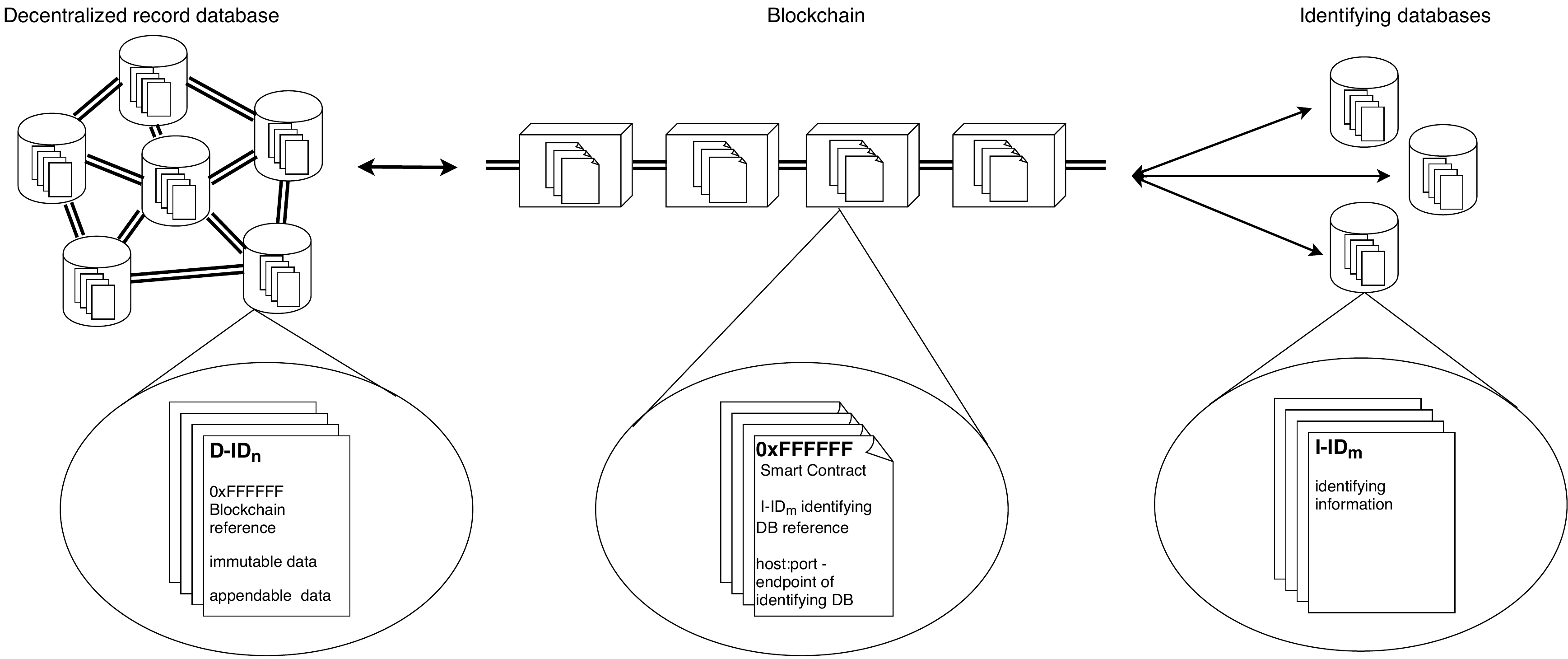}
	\caption{Data Structure. The identifying databases (right) are hosted by the custodians and hold personal information of individuals in single json blob entries, characterized by a unique id. These are linked to user data via smart contracts (center) on the blockchain (one per user). These smart contracts store a url to an identifying database API endpoint as well as the id of the entries holding the personal information in the identifying database. A smart contract is uniquely identifiable by its address.  The decentralized and distributed record database (left) also holds entries in the form of json blobs. They contain a link to the identifying smart contract on the blockchain, as well as appendable and immutable data. Appendable data is the record itself such that it can be updated. Immutable are any links to other data entries, include the pointer to the smart contract.\label{data_structure}}
\end{figure*}

\begin{figure*}
	\centering
	\includegraphics[scale=.45]{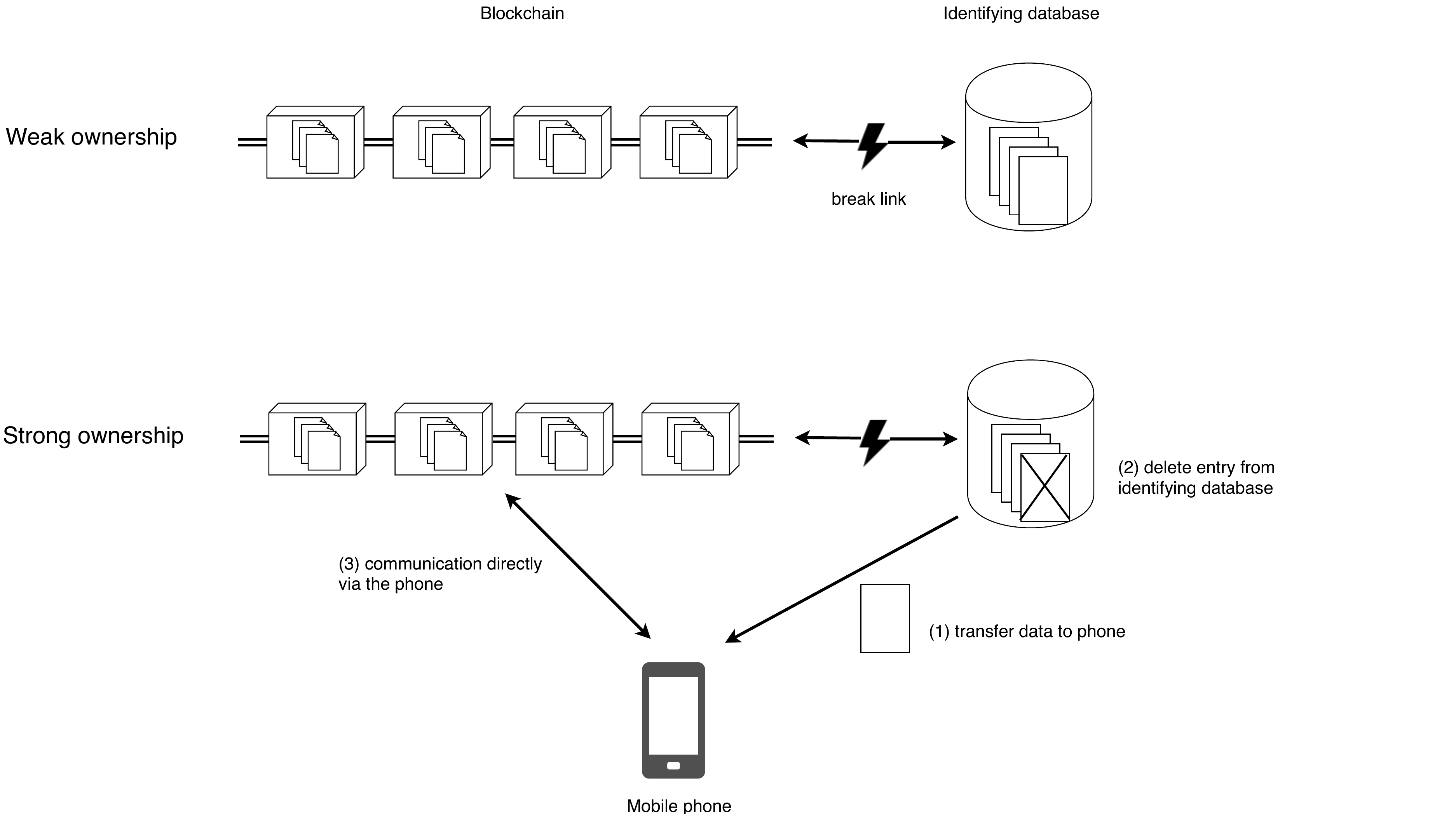}
	\caption{Notion of Ownership. Weak notion of ownership (top) is obtained by scrambling the link between the smart contract on the blockchain (left) and the identifying database holding the personal information (right). This is done by updating the data entry's id in the identifying database without updating it in the smart contract. Strong notion of ownership (bottom) can be obtained with a self-sovereign identity system. Personal information is transfered from the identifying database to the data vault on the mobile phone and deleted from the identifying database. The smart contract is updated to link to the self-sovereign identity, usually a blockchain address. Now, the system will communicate directly with the data owner's phone.\label{ownership}}
\end{figure*}

There are numerous applications for the architecture we introduced, including the storage of academic and medical records. To emphasize the possibilities and advantages our system offers and to explain its functionalities in greater detail, this sections provides a detailed discussion of two use cases. Figure \ref{Fig::Architecture} serves as a visual guide.

\subsection{Academic Records}\label{Sec::UseCases:Academic}
In the application of academic records, the custodians are the universities--or more precisely university administrators and staff--and the data owner is a student. Until today, all student data is stored in a large centralized database at the university, shielded from the outside. There is no possibility for students or third parties to interact with it and therefore the value of the data remains underutilized. In our system, as long as the student has not claimed ownership of her personal data, it will also be stored in a university-hosted centralized CRUD database, the entry being created on registration. However, academic records like marks or degree descriptions are stored in a distributed and decentralized database. Furthermore, the university administrator will deploy a smart contract to the blockchain containing the student's entry id in the identifying database on registration as well as the url to the database endpoint, linking academic records and personal information.

After completion of the first assignment (note that it does not have to be the final course mark but can also be a component of the course), the lecturer or departmental secretary will upload the marks to the consortium of decentralized database nodes, which are run by the universities. A mark entry includes a link to the course description and the degree description, both are entries in the decentralized database, as well as a link to the student's smart contract on the blockchain. There is no identifying information stored because they can be retrieved by looking up the student's smart contract, collecting the id for the identifying database, and finding the relevant information in said database. It is worth emphasizing that only university staff will be permissioned to upload data to the decentralized and distributed database, which is possible because due to our restrictive APIs.

Students are encouraged to claim ownership of their personal data. In the weak form of ownership, it will be as simple as clicking a button in the password protected section to ensure authority. This will generate a QR code the student has to scan with her mobile device and thereby sign into the system with her cryptographic identity. This identity, an address on the blockchain, will become the new owner of the smart contract holding the id of the corresponding entry in the identifying database. Furthermore, the student has the power to exclude anybody from retrieving her personal information by updating the id in the identifying database but not updating it in her smart contract. She can grant read-permission by revealing her new id to requesting third parties. This can be revoked by updating the id again. Universities will most likely not allow the strong notion of ownership, therefore it will be discussed in Section \ref{Sec::UseCases:Medical}.

Third parties in the use case of an academic record storage system are mainly employers and bursary donors but it is not limited to these two groups. Of course, they have to register with the platform first. There are two possible scenarios how they would want to search the data. They either want to search for a variety of students that match certain criteria or they want to check on the progress of one particular student of interest.

In the first case, they send a search request to the decentralized database that queries the data to find matching candidates. Once those have been identified, it will return a list of possible candidates which can not be identified but are anonymized by their smart contract address. The query will further be processed by calling the smart contracts and asking for the students' id in the identifying database and the URL of the database it is stored in. If the student has claimed ownership of the data already, she will be notified and has to decide whether she wants to share her personal information with this particular third party. If she has not claimed ownership yet, the algorithm will look for the id in the identifying database and retrieve the personal information.

The second scenario will be relevant for a bursary donor that wants to check the progress of a particular student she is supporting. If this student has claimed ownership, the donor will have to contact her to request the smart contract address which is used to reverse look-up the academic records. If the student has not claimed ownership yet, the donor can search the identifying database of the corresponding university to find the entry id, which is then used to retrieve the smart contract address by means of a search contract. This search contract stores a mapping from id to smart contract address. Again, the address is the query string to find the relevant data in the decentralized and distributed database.

The system comes with several advantages over the current system. Academic data can be utilized in various beneficial ways. The most obvious one is the match making of students to potential employers or bursary donors from the explanation above. However, the granular data can also be analyzed to send students personal nudges--without knowing who they are--to increase their performance in an upcoming test. If the student has revealed her identity, lecturers and course conveners can also reach out to students at risk of failing a course or dropping out of a degree and support them as best as possible. And even if the student does not reveal her identity, lecturers can contact her anonymously by sending a message via the platform.

\subsection{Medical records}\label{Sec::UseCases:Medical}
The advantage of storing medical data in a system like the proposed one is obvious. All data is stored in one system, accessible from everywhere in case of an emergency. Furthermore, medical research could be conducted on the most accurate data set there is without revealing any identifying information--an important aspect in some jurisdictions.

Again, the record database is a distributed, decentralized database, that could be run by any custodian--physician, clinic, or hospital--that has the ressources to do so. Realistically, these are larger entities that are storing large amounts of data on their inhouse databases at the moment and have the ressources to reallocate. These medical records include diagnoses, test results, treatment plans, pharmaceutical prescriptions, among others. However, they do not contain personal information but only the address of the smart contract representing the individual on the blockchain.

In contrast to the academic records where a student is only registered at one university, patients visit multiple physicians, so there is no possibility to migrate all data concerning one patient at once. Ideally, patients can be identified by their medical insurance number, making sure that every person is only represented by one smart contract and by one entry in the identifying database. However, they may also be the need for a central organization that handles the onboarding of patients onto the system.

Claiming ownership of the data could be accomplished as explained in the example above. Patients log into their personal dashboard, click on a button, and identify themselves using a mobile phone. The address sending that request is made the new owner of the smart contract and now has the right to update the id of their entry in the identifying database but not in the smart contract, thereby revoking access to their personal data and claiming ownership.

Since the centralized databases of identifying information are the most vulnerable link in this architecture, the patient can also decide to obtain strong ownership of her data. Instead of updating the id of her entry in the identifying database, she can move that data into the data vault of her open identity system app, deleting the entry in the identifying database altogether. The blockchain smart contract is updated by assigning the patient's address to the id pointer. This enables the patient to decide on a case-by-case basis whether she wants to share her personal information or not. Every time somebody requests them, she is notified via the app on her phone. In reverse, when the patient visits a doctor, she can reveal her smart contract address and the doctor has access to all medical records corresponding to her. Since a physician is a custodian in the system, she will also have the right to append this medical record.

%
%





%
\section{CONCLUSION}\label{Sec::Conclusion}
%

Currently, the concept of strong data ownership is discussed, but not implemented for university- or medical data. Once a datum has been shared, there is no possibility to prevent the counterpart from copying and sharing or selling it. Consequently, very personal information like student marks or medical records are stored in centralized databases, hidden away from the public and therefore not ideally exploited to improve individuals' lives.

We introduce a system that restores a weak notion of data ownership by leveraging new blockchain technologies like decentralized and distributed ledgers and databases. Data ownership is defined as providing the individual with the possibility to grant and revoke access to her identifying information, which can be used to provide her with tailor-made services. De-identified records are stored in a decentralized and distributed database such that they can be queried by permitted users. Access control is enforced via restrictive APIs and a user interface. Identifying information, and only those, are stored in several centralized databases, hosted by custodians. The data is 
protected by the same access control system as the distributed database. Individual-specific smart contracts on a distributed ledger serve as the link between those two types of databases. They store the URL to the identifying database the record is stored in as well as its id. Decentralized and distributed database will use the smart contracts' address as identifier, therefore de-identifying the data. 

When claiming data ownership, the individual-specific smart contract is signed over to the user. This user now has the ability to prevent others from identifying her by updating the id of her record in the identifying database without updating it in the smart contract. Moreover, she can achieve strong ownership of her data by transferring it from the identifying database to a data vault on her phone using a sovereign-identity service. The smart contract will be updated to point to that vault. This allows the user to decide whether she wants to reveal her identifying information or not on a case-by-case basis.

The current system is not without drawbacks. The weakest spot is the identifying databases that are hosted by custodians. They are not byzantine-fault tolerant and could be easily tampered with. Whenever there is a decentralized and distributed CRUD database available that fulfills the necessary privacy requirements, it can be considered as a substitute. This is left for future iterations of the system.

The architecture we introduce allows for the possibility to draw valuable insights from de-identified data that would otherwise not be accessible to the public. In the use case of academic records, it can be utilized to encourage students to keep up their good work since a national or even international comparison to other students is possible. Furthermore, it can help to find the perfect bursary donor or future employer. If medical records where stored in the above introduced manor, they were accessible everywhere in the world if there was an emergency--no request to the local physician required. Additionally, it will be the most complete medical database available to researches, allowing them to draw live-changing conclusions.

%
%

\end{document}